\documentclass[prl,twocolumn,superscriptaddress,showpacs]{revtex4}

\usepackage{graphicx}
\usepackage{dcolumn}
\usepackage{amsmath}
\usepackage{color}

\newcommand{\etal}{\textit{et al.}}

\begin{document}
\title{Optical evidence for symmetry changes above the N\'{e}el temperature in KCuF$_3$}

\author{J.~Deisenhofer}
\affiliation{D\'{e}partement de Physique de la Mati\`{e}re Condens\'{e}e,
Universit\'{e} de Gen\`{e}ve, CH-1211 Gen\`{e}ve 4, Switzerland}

\author{I.~Leonov}
\affiliation{Abdus Salam International Centre for Theoretical
Physics, Trieste 34014, Italy}

\author{M.~V.~Eremin}
\affiliation{Kazan State University, 420008 Kazan, Russia}

\author{Ch.~Kant}
\affiliation{EP 5, EKM, Institute for Physics, Augsburg University,
D-86135 Augsburg, Germany}

\author{P.~Ghigna}
\affiliation{Dipartimento di Chimica Fisica ``M.~Rolla'',
Universit\`{a} di Pavia, I-27100 Pavia, Italy}

\author{F.~Mayr}
\affiliation{EP 5, EKM, Institute for Physics, Augsburg University,
D-86135 Augsburg, Germany}

\author{V.~V.~Iglamov}
\affiliation{Kazan State University, 420008 Kazan, Russia}

\author{V.I.~Anisimov}
\affiliation{Institute of Metal Physics, 620219 Yekaterinburg
GSP-170, Russia}

\author{D.~van der Marel}
\affiliation{D\'{e}partement de Physique de la Mati\`{e}re Condens\'{e}e,
Universit\'{e} de Gen\`{e}ve, CH-1211 Gen\`{e}ve 4, Switzerland}

\date{\today}

\begin{abstract}
We report on optical measurements of the 1D Heisenberg
antiferromagnet KCuF$_3$. The crystal-field excitations of the
Cu$^{2+}$ ions have been observed and their temperature dependence
can be understood in terms of magnetic and exchange-induced dipole
mechanisms and vibronic interactions. Above $T_N$ we observe a new
temperature scale $T_S$ characterized by the emergence of narrow
absorption features that correlate with changes of the orbital
ordering as observed by Paolasini \etal~[Phys. Rev. Lett.
\textbf{88}, 106403 (2002)]. The appearance of these optical
transitions provides evidence for a symmetry change above the Neel
temperature that affects the orbital ordering and paves the way for
the antiferromagnetic ordering.

\end{abstract}



\maketitle

The charge-transfer insulator KCuF$_3$ is a prototype material for a
cooperative Jahn-Teller (JT) effect, orbital ordering (OO), and
low-dimensional magnetism \cite{Kadota67}. KCuF$_3$ stands in line
with LaMnO$_3$, the parent compound of the colossal
magnetoresistance manganites, as role models for orbitally ordered
compounds, where collective orbital modes may occur in the
excitation spectrum \cite{Saitoh01,Ruckamp05}. Moreover, KCuF$_3$ is
one of the rare examples of an ideal one-dimensional (1D)
antiferromagnetic (AFM) Heisenberg chain. This is related to the
particular OO in KCuF$_3$, in which a single hole alternately
occupies $3d_{x^2-z^2}$ and $3d_{y^2-z^2}$ orbital states of the
Cu$^{2+}$ ions with a $3d^9$ electronic configuration
\cite{Kugel82}. The cooperative JT distortion is characterized by
CuF$_6$ octahedra elongated along the $a$ and $b$ axis and arranged
in an antiferrodistortive pattern in the $ab$-plane. A crossover
from a 1D to a three-dimensional behavior occurs at the N\'{e}el
temperature $T_N$ =39~K resulting in a long-range ($A$-type) AFM
ordering, but the fingerprint of a 1D spin chain, the spinon
excitation continuum, is still observable below $T_N$ and persists
up to about 200~K \cite{Lake05}. Together with a strong suppression
of the ordered moment found below $T_N$, this establishes the impact
of quantum fluctuations on the magnetism of KCuF$_3$ \cite{Lake00}.

To date even the room temperature (RT) crystal structure of KCuF$_3$
seems not to be determined unambiguously. The original assignment of
tetragonal symmetry \cite{Okazaki61} was challenged by the claim of
an orthorhombic structure \cite{Hidaka98}. Although such a
distortion seems to allow for a better understanding of Raman
\cite{Ueda91} and electron paramagnetic resonance (EPR) properties
\cite{Yamada89}, discrepancies remain when trying to reconcile the
observation of AFM resonances within the proposed symmetries
\cite{Li05}. Paolasini and coworkers reported an X-ray resonant
scattering (XRS) study of the OO-superlattice reflection in KCuF$_3$
and found a considerable increase of its intensity above $T_N$,
indicating changes of the lattice degrees of freedom prior to
magnetic ordering \cite{Paolasini02,Caciuffo02}.

Here we report optical spectra of the crystal-field (CF) excitations
in KCuF$_3$ in the near infrared range. The observed energy
splittings are compared to theoretical estimates using crystal-field
theory (CFT) and to a direct calculation in the framework of the
LDA+$U$ approximation \cite{Streltsov05}. Furthermore, we discovered
the appearance of a very sharp fine structure related to the $d$-$d$
absorption bands below $T_S$=50~K, which is interpreted as a new
temperature scale above $T_N$, where a dynamic distortion of the
environment of the Cu$^{2+}$ ions becomes static.

The single crystals (see Ref.~\onlinecite{Paolasini02} for details
on crystal growth) were oriented by Laue diffraction and cut along
the (110)-plane. Polarization-dependent transmission measurements
were performed using a BRUKER IFS 66v/S Fourier-transform
spectrometer with a $^4$He cryostat (OXFORD) in the frequency range
between 5000-16000 cm$^{-1}$ and for temperatures from 4-300~K. The
absorption coefficient $\alpha$ was obtained using the standard
relation for a semitransparent layer \cite{Dressel} and
$\sigma_1$=$\alpha nc\epsilon_0$ was obtained by estimating the
refractive index $n$ from $\epsilon_1$$\approx$ 2.2 given by
ellipsometry data at RT. This value is in good agreement with
$\epsilon_1$$\approx$ 2.3 obtained theoretically \cite{Nikiforov96}.

\begin{figure}[t]
\centering
\includegraphics[width=80mm,clip]{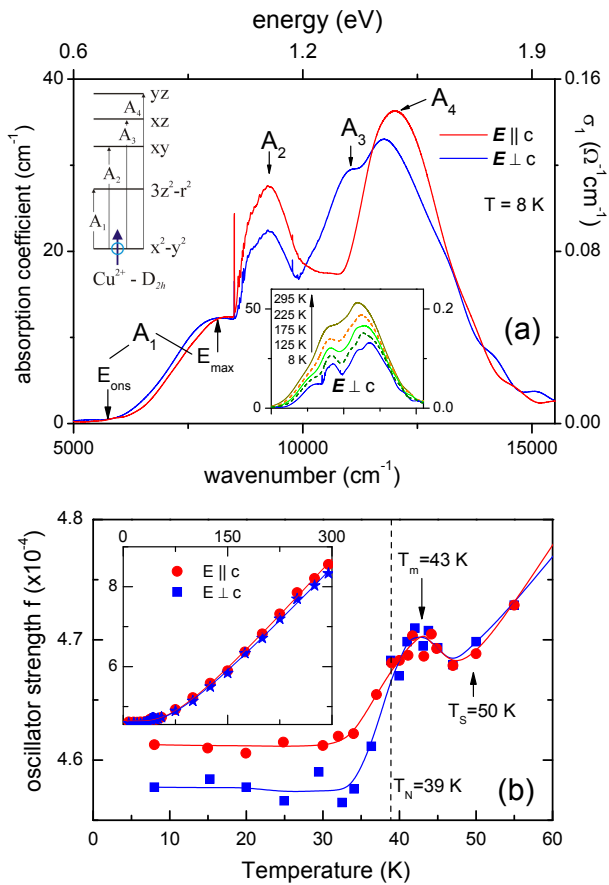}
\vspace{2mm} \caption[]{\label{AbsCoeff4k} (a) Absorption spectra
for $E\parallel c$ and $E\perp c$ at 8~K. Arrows indicate the four
absorption bands A$_1$--A$_4$ related to the CF splitting for
Cu$^{2+}$ in $D_{2h} symmetry$. For A$_1$ the onset energy $E_{ons}$
and the maximum at $E_{max}$ are indicated. Inset: Evolution of
absorption spectra with temperature for $E\perp c$. (b) Anomalies of
the oscillator strength $f$ above $T_N$. Lines are guides to the
eye. Inset: Full temperature dependence of $f$, solid lines are fits
described in the text.}
\end{figure}

\begin{figure}[t]
\centering
\includegraphics[width=80mm,clip]{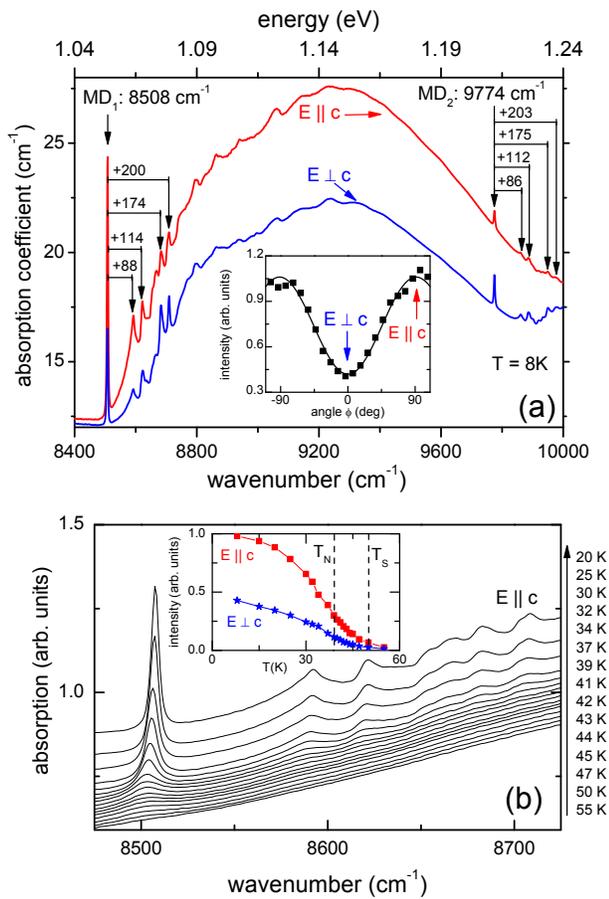}
\vspace{2mm} \caption[]{\label{Tdepvibron1} (a) Fine structures
related to $A_2$ and $A_3$ for $E\parallel c$ and $E\perp c$ at 8~K.
Frequency shifts with respect to MD$_1$ and MD$_2$ are indicated in
cm$^{-1}$. Inset: Polarization dependence of the intensity of MD$_1$
and a fit $\propto \sin^2\phi$. (b) Emerging of the fine structure
peaks below $T_S=50$~K for $E\parallel c$. Inset: Temperature
dependence of the intensity of MD$_1$ for both orientations. Lines
are guides to the eyes.}
\end{figure}

In Fig.~\ref{AbsCoeff4k}(a) we show the absorption coefficient
$\alpha$ for light polarized parallel ($\mathbf{E}\parallel c$) and
perpendicular ($\mathbf{E}\perp c$) to the antiferromagnetically
coupled $c$-axis at 8~K. For both orientations one can identify four
broad excitation bands which are indicated by arrows and called
A$_1$--A$_4$ in the following. The onset and maximum energies of
these bands are given in Table \ref{table1}. One can recognize
rather symmetric lineshapes for A$_1$ and A$_4$, while A$_2$
exhibits a very asymmetric lineshape with a steep absorption edge at
the low-energy side. Since A$_3$ is visible mainly as a shoulder in
the spectrum, its lineshape is not evident. Moreover, a sharp
spike-like feature can be seen at the onset of the A$_2$ band (see
Fig.~\ref{Tdepvibron1}(a) for an enlarged scale), which will be
discussed in detail below.

Comparing the two spectra in Fig.~\ref{AbsCoeff4k}(a) one finds that
the intensities of A$_2$ and A$_4$ are enhanced for
$\mathbf{E}\parallel c$, while A$_3$ seems to be much more intense
for $\mathbf{E}\perp c$. The temperature behavior of the absorption
bands is illustrated in the inset of Fig.~\ref{AbsCoeff4k}(a) for
$\mathbf{E}\perp c$. The bands broaden and gain intensity with
increasing temperature. The nature of these absorption bands can be
understood in terms of local $d$-$d$ CF excitations of the Cu$^{2+}$
ions in a distorted octahedral environment \cite{footnote}. Usually,
electric dipole (ED) transitions between $d$-levels are
parity-forbidden, but they can become allowed as a consequence of a
perturbation of the CF potential by phonons. A fingerprint of such
\textit{vibronically} allowed transitions is the temperature
dependence of their oscillator strength $f$ given by
\begin{eqnarray}
\int_{\omega_1}^{\omega_2}\sigma_1(\omega)d\omega=\frac{\pi}{2}\frac{e^2N}{m}f,
\end{eqnarray}
where $N$ is the number of absorption centers per unit volume
\cite{Sugano}. Assuming that only a single vibrational mode with
frequency $\omega_{vib}$ contributes to the intensity of the
absorption one expects $f=f_{vib}\coth (\hbar\omega_{vib}/2k_BT)$
\cite{Sugano,Hitchman79}. By adding a temperature independent term
$f_{mag}$ one usually accounts for the contribution of magnetic
dipole (MD) transitions. The order of magnitude of
$10^{-3}$-$10^{-4}$ is in agreement with theoretical expectations
\cite{Sugano}.

We show as an inset in Fig.~\ref{AbsCoeff4k}(b) the temperature
dependence of $f$ obtained by integration between
$\omega_1$=5000~cm$^{-1}$ and $\omega_2$=16000~cm$^{-1}$ for both
orientations. Fitted over the whole temperature range from 8-295~K
$f$ appears to be well described (solid lines) by this approach with
$\omega_{vib}$=148~cm$^{-1}$, 144 cm$^{-1}$,
$f_{vib}$=2.1$\times10^{-4}$, 1.9$\times 10^{-4}$, and
$f_{mag}$=2.5$\times10^{-4}$, 2.7$\times 10^{-4}$ for
$\mathbf{E}\parallel c$ and $\mathbf{E}\perp c$, respectively. When
zooming in to the low-temperature regime in the vicinity of $T_N$ as
shown in Fig.~\ref{AbsCoeff4k}(b), one can clearly see deviations
from this behavior starting at below a temperature $T_S$$\approx$
50~K. An additional contribution to $f$ leads to a maximum at around
43~K and a subsequent decrease upon AFM ordering at 39~K towards a
constant low-temperature value. Note that the increase below 50~K is
directly related with the appearance of the sharp-absorption
features at the broad bands $A_2$ and $A_3$ (see
Fig.~\ref{Tdepvibron1}(a)).

\begin{table}[b]
\caption[]{\label{table1}Comparison of $E_{ons}$ and $E_{max}$ for
the energies of the onsets and the maxima of bands A$_1$--A$_4$ with
theoretical estimates $E_{CFT}$, $E_{\rm LDA}$ and $E_{{\rm LDA}+U}$
obtained as described in the text. The energies are given in eV.}
\quad

\begin{tabular}{c|c|c||c|c|c}
  \hline
  \hline
                & \,\,$E_{ons}$\, \, &\, $E_{CFT}$ \, &\, $E_{max}$ \, &\, \,$E_{{\rm LDA}}$\, \,& \,$E_{{\rm LDA}+U}$\,  \\
  \hline
  $yz$          &      1.34     &     1.34       &   1.46      &  1.62  &     1.50  \\
  $xz$          &      1.21     &     1.26       &   1.37      &  1.37  &     1.41  \\
  $xy$          &      1.05     &     1.06       &   1.15      &  1.28  &     1.31  \\
  $3z^2-r^2$    &      0.71     &     0.69       &   1.02      &  0.94  &     1.03  \\
  $x^2-y^2$     &        0      &       0        &      0      &  0     &     0     \\
  \hline
  \hline

\end{tabular}
\end{table}

Considering the space group $D_{4h}^{18}$-$I4/mcm$ \cite{Okazaki61}
and the local $D_{2h}$-symmetry at the Cu$^{2+}$ site , a complete
splitting of the CF levels is expected [see
Fig.~\ref{AbsCoeff4k}(a)]. We compare the experimental splittings to
theoretical estimates via two different approaches in
Tab.~\ref{table1}. The onset energies $E_{ons}$ of $A_1$-$A_4$
correspond to the purely electronic splittings and are estimated by
$E_{CFT}$ obtained using the exchange charge model in CFT
\cite{Malkin77,Malkin87}. The calculation describes the experimental
splittings nicely and further parameters like anisotropic $g$-values
and the nuclear quadrupole resonance properties were derived, in
good agreement with experimental data \cite{Yamada89,Mazzoli}.

In the second approach, the electronic structure of KCuF$_3$ was
calculated self-consistently using LDA(+$U$). According to previous
LDA calculations \cite{Liechtenstein95,Binggeli04} the Coulomb and
exchange interaction parameters $U$ and $J$ were taken to be 8.0 and
0.9~eV, respectively. Following Streltsov {\it et al.}
\cite{Streltsov05}, we additionally estimate $E_{{\rm LDA}+U}$ as
the difference between the ground state energy and an energy of an
excited state in which the hole is artificially put on one of the
higher levels by applying constrained LDA+$U$ calculations. This
constrained method takes into account the relaxation of the
electronic system, whereas the lattice remains frozen obeying the
Franck-Condon principle. Hence, the resulting CF splittings have to
be compared with $E_{max}$ of the absorption bands in
Tab.~\ref{table1}. Both calculations describe the experimental
findings, but including electronic relaxation gives an overall
better agreement between theory and experiment.

Having described the origin of the broad band spectra on a
quantitative level, we now concentrate on the observation of two
sets of sharp absorption features highlighted in
Fig.~\ref{Tdepvibron1}(a), one indicating the onset of A$_2$ at
8508~cm$^{-1}$ followed by a whole sideband of excitations, and a
second set starting at 9775~cm$^{-1}$ at the anticipated onset of
A$_3$. Both sets are visible for $\mathbf{E}\parallel c$ and
$\mathbf{E}\perp c$. As an inset of Fig.~\ref{Tdepvibron1}(a) we
show the polarization dependence of the intensity of the peak at
8508~cm$^{-1}$ following a $\sin^2\phi$ dependence, with $\phi$
being the angle between the polarization and the crystal axes. While
the latter peak is more intense for $\mathbf{E}\parallel c$, the
peak at 9775~cm$^{-1}$ is more intense for $\mathbf{E}\perp c$. Note
that $H_z$ in the local coordinate system corresponds to
$\mathbf{B}\perp c$ or $\mathbf{E}\parallel c$ and $H_y$ corresponds
to $\mathbf{B}\parallel c$ or $\mathbf{E}\perp c$, with
$\mathbf{B}=\mu(T) \mathbf{H}$, where $\mu(T)$ is the permeability
which is usually assumed to be one in optical experiments, but can
increase significantly upon magnetic ordering. The observed
intensity maxima comply with the selection rules for MD transitions.

Concerning the side bands of both MD transitions, we want to
emphasize that the strongest features appear at a distance of
88(86), 114(112), 174(175), and 200(203)~cm$^{-1}$ from the
MD$_1$(MD$_2$) transitions [see Fig.~\ref{Tdepvibron1}(a)]. To
understand the origin of the whole entity of MD transitions and
their sidebands, we have to consider their temperature dependence,
which is displayed in Fig.~\ref{Tdepvibron1}(b) for
$\mathbf{E}\parallel c$. The fine structure emerges only below a
temperature $T_S$=50~K, the temperature where the overall oscillator
strength starts to deviate from a purely vibrational behavior with a
constant contribution from MD transitions
(Fig.~\ref{AbsCoeff4k}(a)). The integrated intensity of the MD
transition at 8508~cm$^{-1}$ as a function of temperature is shown
as an inset in Fig.~\ref{Tdepvibron1}(b) for both polarizations
without any anomaly visible at $T_N$=39~K.

It was shown that magnons can become optically active via an
exchange-induced dipole mechanism \cite{Greene65,Tanabe67}, with the
optical excitation energy corresponding to the magnon energy at the
Brillouin zone boundary. In KCuF$_3$, neutron scattering
investigations reported zone boundary energies of magnon excitations
at 88.7, 117, and 221~cm$^{-1}$ \cite{Satija80,Lake00}. The first
two magnon energies correspond very nicely with the first and second
sideband peaks, the third sideband peak is in agreement with a
two-magnon process involving the 88.7~cm$^{-1}$-magnon, the fourth
would correspond to two-magnon process where both the 88.7 and the
117~cm$^{-1}$ magnon participate. For the 221~cm$^{-1}$-magnon we
can not resolve any sideband in absorption. In this scenario the
temperature dependence of the MD transitions and the magnon
sidebands below $T_S$ is attributed to the increase of $\mu(T)$ in
the vicinity of AFM ordering and can gain additional intensity due
to the local spin-spin expectation value $\langle
\mathbf{S}_{i}(\mathbf{S}_{i-1}+\mathbf{S}_{i+1})\rangle$ which
contributes to the exchange-induced dipole moment $\sum_j
\mathbf{\Pi}_{ij}(\mathbf{S}_i\mathbf{S}_j)$ for the Cu ion at site
$i$. For $\mathbf{E}\parallel c$ the spin-spin correlations in the
$ab$-plane are important, for $\mathbf{E}\perp c$ the contributing
correlations are the ones along the $c$-axis \cite{Tanabe67}.

Here we would like to point out that an anomalous behavior in
birefringence \cite{Iio78} and neutron scattering studies
\cite{Tennant95} has been reported at around $T_m$=43~K, exactly the
temperature where $f$ exhibits a maximum (Fig.~\ref{AbsCoeff4k}(b)).
At the same temperature a significant change of the OO superlattice
reflection at the Cu $K$-edge has been reported by a XRS study
\cite{Paolasini02,Caciuffo02}. Since XRS at the $K$-edge probes the
JT distortion induced by the OO \cite{Benedetti01}, the anomaly
certainly involves a change in the lattice degrees of freedom. On
account of LDA+$U$ calculations it has previously been suggested
that the anomalous XRS data might indicate a structural transition
\cite{Binggeli04}.

In literature we find two hints concerning a symmetry lowering in
KCuF$_3$, a X-ray diffraction study at RT claiming the presence of
an orthorhombic distortion due to a shift of the apical F$^-$ ions
connecting the CuF$_6$ octahedra along the $c$-axis \cite{Hidaka98},
and a Raman study of the phonon modes at 10~K, observing the
splitting of two $E_g$ Raman modes \cite{Ueda91}. To get a more
consistent picture of what happens at $T_S$=50~K, the temperature
dependence of the split Raman modes was investigated and the modes
were found to soften strongly with decreasing temperature and split
only below $T_S$=50~K \cite{Lemmens08}. The occurrence of this
splitting $T_S$=50~K is a clear sign for a reduced (static) lattice
symmetry. Moreover, a recent approach allows to describe the EPR
properties by assuming that the displacement of the apical F$^-$
ions away from the $c$-axis is dynamic in nature, but becomes static
at $T\approx T_N$ \cite{Eremin08}. The phonon modes contributing to
the oscillation of the F$^-$ ions should soften with decreasing
temperature as the displacement of the F ions freezes in and becomes
static at $T_S$=50~K, in agreement with the Raman data. This static
distortion manifests itself in an increase of intensity of the MD
transitions in optics, a splitting of Raman-active phonons, and the
increase in intensity of the orbital-ordering superstructure
reflection in XRS (lattice fluctuations produce orbital fluctuations
which might reduce the XRS intensity above 43~K). The freezing-in of
the distortion leads to an additional magnetic anisotropy via a
Dzyaloshinsky-Moriya interaction, which favors the alignment of the
spins and allows for the AFM transition to take place at $T_N$=39~K.
Next-nearest-neighbor spin correlations develop already below $T_S$
leading to the observation of optically-active magnons via the
exchange-induced dipole mechanism.

To summarize, we observed broad phonon-assisted CF excitations of
the Cu$^{2+}$ ions with a pronounced fine structure below
$T_S$=50~K. These sharp absorption peaks are identified as magnetic
dipole transitions and optically active magnons gaining intensity
due a exchange-induced dipole moment. We argue that at $T_S$=50~K a
dynamic displacement of F$^-$ ions becomes static and leads to
symmetry change of the lattice. Given the fact that, e.g.\;
manganites reportedly exhibit changes of the OO superlattice
reflection as measured by XRS in the vicinity of $T_N$
\cite{Murakami98a} and softening of Raman-active phonon modes
\cite{Choi08}, too, one may speculate that the phenomena described
in the present paper could be present in many AFM transition metal
compounds.

\begin{acknowledgments}
It is a pleasure to thank N.P.\;Armitage, F.\;Banfi, N.\;Binggeli,
M.\;Gr\"{u}ninger, A.B.\;Kuzmenko, C.\;Mazzoli, A.\;Loidl and
D.\;Zakharov for fruitful discussions. We acknowledge partial
support by the BMBF via contract number VDI/EKM 13N6917, by the DFG
via SFB 484, and by the Swiss NSF through NCCR MaNEP.
\end{acknowledgments}


\begin{thebibliography}{39}
\expandafter\ifx\csname
natexlab\endcsname\relax\def\natexlab#1{#1}\fi
\expandafter\ifx\csname bibnamefont\endcsname\relax
  \def\bibnamefont#1{#1}\fi
\expandafter\ifx\csname bibfnamefont\endcsname\relax
  \def\bibfnamefont#1{#1}\fi
\expandafter\ifx\csname citenamefont\endcsname\relax
  \def\citenamefont#1{#1}\fi
\expandafter\ifx\csname url\endcsname\relax
  \def\url#1{\texttt{#1}}\fi
\expandafter\ifx\csname
urlprefix\endcsname\relax\def\urlprefix{URL }\fi
\providecommand{\bibinfo}[2]{#2}
\providecommand{\eprint}[2][]{\url{#2}}


\bibitem{Kadota67} S.~Kadota \etal, J.~Phys.~Soc.~Jpn.
{\bf 23}, 751 (1967).

\bibitem{Saitoh01}
E.~Saitoh \etal, Nature {\bf 410}, 180 (2001).

\bibitem{Ruckamp05} R. Ruckamp \etal, New J.~Phys.~\textbf{7}, 144 (2005).


\bibitem{Kugel82} K.I.~Kugel and D.I.~Khomskii, Sov.~Phys.~Usp.~{\bf 25}, 231
(1982).


\bibitem{Lake05} B.~Lake \etal, Nature Materials~{\bf 4}, 329 (2005).

\bibitem{Lake00} B.~Lake \etal, Phys.~Rev.~Lett. {\bf 85}, 832 (2000).

\bibitem{Okazaki61} A.~Okazaki and Y. Suemune, J.~Phys.~Soc.~Jpn.~{\bf 16}, 176 (1961).

\bibitem{Hidaka98} M.~Hidaka \etal, J.~Phys.~Soc.~Jpn.
{\bf 67}, 2488 (1998).


\bibitem{Ueda91} T.~Ueda \etal, Solid~State~Comm. {\bf 80}, 801 (1991).

\bibitem{Yamada89} I.~Yamada \etal, J.~Phys.:~Cond.~Mat. {\bf 1}, 3397 (1989).

\bibitem{Li05} L.~Li \etal, J.~Phys.:~Cond.~Mat. {\bf 17}, 2749 (2005).


\bibitem{Paolasini02} L.~Paolasini \etal, Phys.~Rev.~Lett. {\bf 88}, 106403 (2002).

\bibitem{Caciuffo02} R.~Caciuffo \etal, Phys.~Rev.~B \textbf{65}, 174425
(2002).

\bibitem{Streltsov05} S.V. Streltsov \etal, Phys.~Rev.~B {\bf 71},
245114 (2005).

\bibitem{Dressel} M. Dressel and G. Gr\"{u}ner, \textit{Electrodynamics of Solids}, Cambridge University Press, Cambridge, (2002).

\bibitem{Nikiforov96} A.E.~Nikiforov and S.Yu.~Shashkin, Phys. Solid State~{\bf 38}, 1880 (1996).

\bibitem{footnote} The local $z$-direction is along the longest Cu-F bond resulting in a
$x^2-y^2$-orbital ground state.


\bibitem{Sugano} S.~Sugano \etal. \textit{Multiplets of Transition Metal Ions in Crystals},
 Academic Press, London, (1970).

\bibitem{Hitchman79} M.A. Hitchman and P.J. Cassidy, Inorg. Chem. \textbf{18}, 1745
(1979).


\bibitem{Malkin77} M.V. Eremin \etal, phys. stat. solid.(b) \textbf{79}, 775
(1977).

\bibitem{Malkin87} B.Z.~Malkin. \textit{Spectroscopy of solids containing rare-earth ions},
ed. A.H.Kaplyanskii and R.M. Macfarlane, Elsevier (1987).

\bibitem{Mazzoli} C. Mazzoli \etal, J. Magn. Magn. Mater. \textbf{242-245}, 935 (2002); \textit{ibid}. \textbf{272-276}, 106 (2004).

\bibitem{Liechtenstein95} A.I.~Liechtenstein \etal,
Phys.~Rev.~B {\bf 52}, 5467 (1995).

\bibitem{Binggeli04} N.~Binggeli and M.~Altarelli, Phys.~Rev.~B \textbf{70}, 085117 (2004).


\bibitem{Benedetti01} P.~Benedetti \etal, Phys.~Rev.~B {\bf 63}, 060408 (2001).

\bibitem{Iio78} K.~Iio \etal, J.~Phys.~Soc.~Jpn.~{\bf 44}, 1393 (1978).



\bibitem{Tennant95} D.A. Tennant \etal,
Phys.~Rev.~B {\bf 52}, 13381 (1995).


\bibitem{Greene65}
R.L. Greene \etal, Phys.~Rev.~Lett.~{\bf 15}, 656 (1965).

\bibitem{Tanabe67} Y.~Tanabe and K. Gondaira, J.~Phys.~Soc.~Jpn.~{\bf 22}, 573 (1967).


\bibitem{Satija80} S.K.~Satija \etal,
Phys.~Rev.~B {\bf 21}, 2001 (1980).


\bibitem{Lemmens08} V.~Gnezdilov \etal, unpublished.


\bibitem{Eremin08} M.V. Eremin \etal, condmat-0807.3641, unpublished
(2008).


\bibitem{Murakami98a}
Y.~Murakami \etal, Phys.~Rev.~Lett.~{\bf 80}, 1932 (1998),
\textit{ibid}.~{\bf 81}, 582 (1998).

\bibitem{Choi08} K.-Y.~Choi \etal, Phys.~Rev.~B {\bf 77}, 064415 (2008).



\end{thebibliography}


\end{document}